\documentclass[twocolumn,pra, superscriptaddress]{revtex4}
\newcommand{\pagenumbaa}{1}
\usepackage{graphicx}
\usepackage{amssymb}
\usepackage{amstext}
\usepackage{epsfig}
\usepackage{algpseudocode}
\usepackage{amscd}
\usepackage{algorithm}
\usepackage{tikz}

\usepackage{amsthm}
\usepackage{amsmath}
\usepackage{amsfonts}
\usepackage{amssymb,amstext}

\theoremstyle{plain}

\theoremstyle{definition}

\newcommand{\beq}{\begin{equation}}
\newcommand{\eeq}{\end{equation}}

\newcommand{\beqa}{\begin{eqnarray}}
\newcommand{\eeqa}{\end{eqnarray}}

\newcommand{\bal}{\begin{align}}
\newcommand{\eal}{\end{align}}

\newcommand{\bsp}{\begin{equation}\begin{split}}
\newcommand{\esp}{\end{split}\end{equation}}

\newcommand{\bit}{\begin{itemize}}
\newcommand{\eit}{\end{itemize}}

\newcommand{\ben}{\begin{enumerate}}
\newcommand{\een}{\end{enumerate}}

\newcommand{\nn}{\nonumber}

\renewcommand{\sp}[2]{\langle #1,#2 \rangle}
\newcommand{\AR}{\mathbb{R}}

\newcommand*{\ket}[1]{| #1 \rangle}
\newcommand*{\bra}[1]{\langle #1 |}

\newcommand{\HR}{\mathcal{H}}

\newcommand{\id}{\mathbb{I} }

\newcommand{\tr}{\mathrm{tr}}

\newcommand*\circled[1]{\tikz[baseline=(char.base)]{
            \node[shape=circle,draw,inner sep=1pt] (char) {#1};}}

\algnewcommand\algorithmicswitch{\textbf{switch}}
\algnewcommand\algorithmiccase{\textbf{case}}
\algnewcommand\algorithmicassert{\texttt{assert}}
\algnewcommand\Assert[1]{\State \algorithmicassert(#1)}%
\algdef{SE}[SWITCH]{Switch}{EndSwitch}[1]{\algorithmicswitch\ #1\ \algorithmicdo}{\algorithmicend\ \algorithmicswitch}%
\algdef{SE}[CASE]{Case}{EndCase}[1]{\algorithmiccase\ #1}{\algorithmicend\ \algorithmiccase}%
\algtext*{EndSwitch}%
\algtext*{EndCase}%
\begin{document}

%%%%%%%%%%%%%%%%%%%%%%%%%%%%%%%%%%%%%%%%%%%%%%%%%%%%%%%%%%%%%%%%%%%%
% Place your title here
%%%%%%%%%%%%%%%%%%%%%%%%%%%%%%%%%%%%%%%%%%%%%%%%%%%%%%%%%%%%%%%%%%%%
\title{Simultaneous Estimation of Dimension, States and Measurements:\\ Computation of representative density matrices and POVMs}

%%%%%%%%%%%%%%%%%%%%%%%%%%%%%%%%%%%%%%%%%%%%%%%%%%%%%%%%%%%%%%%%%%%%
% Place names of authors here
%%%%%%%%%%%%%%%%%%%%%%%%%%%%%%%%%%%%%%%%%%%%%%%%%%%%%%%%%%%%%%%%%%%%

\author{Cyril Stark}

\affiliation
{Institute for Theoretical Physics, ETH Z\"urich, 8093 Z\"urich, Switzerland}

%%%%%%%%%%%%%%%%%%%%%%%%%%%%%%%%%%%%%%%%%%%%%%%%%%%%%%%%%%%%%%%%%%%%
% Abstract to the abstract
%%%%%%%%%%%%%%%%%%%%%%%%%%%%%%%%%%%%%%%%%%%%%%%%%%%%%%%%%%%%%%%%%%%%

\begin{abstract}

This investigation continues a program aiming at obtaining effective quantum models to describe measurement statistics. In~[Stark, arXiv:1209.5737 (2012)], we have described how the Gram matrix $G$ associated to the prepared states and the performed POVM elements can be estimated via convex relaxation of a rank minimization problem. This Gram matrix $G$ determines the density matrices and the POVM elements uniquely up to simultaneous rotations (with respect to the Hilbert-Schmidt inner product) in the vector space of Hermitian matrices. However, when the description of the experiment needs to be connected to textbook quantum mechanics, explicit expressions for the states and the POVM elements in terms of positive semidefinite matrices are required. In this paper, we describe a heuristic algorithm that takes the state-meaurement Gram matrix $G$ as input and searches explicit realizations of $G$ in terms of quantum mechanically valid density matrices and POVM elements. 

\end{abstract}

\maketitle

\setcounter{page}{\pagenumbaa}
\thispagestyle{plain}

Assume you have built up an experiment that allows you to prepare a quantum system in different states, and subsequently perform different measurements on this system. The catch is that you do not know exactly what states are actually prepared, and what the measurement devices truly measure. But maybe you are willing to assume that the experiment can be repeated such that one can gather statistics. Then, the question is what can be deduced from this knowledge. We have previously introduced a general program for simultaneous estimation of states and measurements. In \cite{Stark2012GramEstimations} we described how the Gram matrix, $G$, associated to the prepared states and the POVM elements can be estimated via convex relaxation of a rank minimization problem. We showed that $G$ is not always uniquely determined by the acquired measurement data.  In \cite{Stark2012RigidityConsiderations} we introduced a method to test whether or not specific assumptions on the Gram matrix can guarantee its uniqueness. The Gram matrix $G$ only determines the density matrices and the POVM elements uniquely up to simultaneous rotations (with respect to the Hilbert-Schmidt inner product) in the vector space of Hermitian matrices. If we compute them via standard factorizations of $G$, we typically do not get matrices which are positive semidefinite as we normally would expect of proper density operators and POVM elements. In this paper we complete the proof of principle of the simultaneous estimation of dimension, states and measurements, in terms of a heuristic algorithm that delivers a collection of positive and normalized density operators for the states, and positive POVM elements for the measurements.

We observe a growing interest in connection with inadequate quantum modeling: In~\cite{MoroderBlattSystematicErrors2012} the authors introduce a method which allow for the falsification of inaccurate modelling of the performed POVMs. Temporal drifts in the measurement setup are discussed in~\cite{FlammiaGross2012}. The papers~\cite{Scarani2011FalseBellViolation, AcinGisinMasanes2006,RossetGisin2012, WoodheadPironio2012} investigate consequences of inappropriate theoretical models. In our work, dimension estimation is tied to the estimation of the states and the measurements. In~\cite{GallergoAcin2010, GallergoAcin2012Nature, GallergoAcin2012} on the other hand, dimension is estimated via linear witnesses.

This paper is organized as follows. In Section~\ref{Sec:setting} we introduce the general setting, define the general task more precisely, and clarify its role in the general program outlined in \cite{Stark2012GramEstimations} and \cite{Stark2012RigidityConsiderations}. 
In Section~\ref{Sec:algorithm} we discuss the main ideas behind our algorithm, and introduce its different subroutines. Section \ref{Sec:switching} considers the necessary `monitoring' of the evolution of the computation, and how this is used for the switching between the different subroutines. In Section~\ref{Sec:examples} we provide some numerical examples to illustrate the application of this technique.

%%%%%
\section{Setting}\label{Sec:setting}

We imagine that we are able to prepare $W$ different states $( \rho_{w} )_{w}$ and to measure $V$ choices of POVMs $(E_{vk})_{vk}$ ($k$ enumerates the different outcomes). To keep the notation simple, we assume that the number of outcomes, $K$, is the same for all measurements. Both the density matrices and the POVMs are unknown a priori even though you might have an educated guess what they might look like. Performing independent repetitions of each of the measurements on each of the states, we can determine frequency distributions $(f_{w,vk})_{k}$ for each pairing of a state `$w$' with a measurement `$v$'. Here, $f_{w,vk}$ denotes the frequency for measuring `$k$' in case we have prepared state `$w$' and measured POVM `$v$'. In the asymptotic limit (i.e., the number of measurement repetitions goes to infinity), $f_{w,vk}$ is equal to its associated probability $p_{w,vk}$. Consequently, 
\[
	f_{w,vk} = \tr(E_{vk} \rho_{w}) = \sum_{i,j=1}^d (E_{vk})_{ij} (\rho_{w})_{ij} = (\vec{E}_{vk})^T \vec{\rho}_{w}.
\]
Here, we have created the vectors $\vec{E}_{vk}$ and $\vec{\rho}_{w}$ in $\mathbb{C}^{d^2}$ out of  the matrices $E_{vk}, \rho_{w}$ by stacking all the matrix-columns on top of each other. When collecting all the frequencies within one data table
\beq\label{def.data.D}
	\mathcal{D} =   \left( \begin{array}{ccc}  f_{1,(1,1)} & \cdots & f_{1,(V,K)} \\  \vdots &   &  \vdots  \\   f_{W,(1,1)} & \cdots  & f_{W,(V,K)}  \end{array} \right),
\eeq
and all the vectors $\vec{\rho}_{w}$ and $\vec{E}_{vk}$ within one matrix
\beq\label{reminder.def.matrix.carrying.st.and.m}
	P = ( \vec{\rho}_{1} \, | \cdots | \, \vec{\rho}_{W} \, | \, \vec{E}_{11}  \, | \cdots |  \, \vec{E}_{VK}  ),
\eeq
we observe that $\mathcal{D}$ appears as off-diagonal blocks in the state-measurement Gram matrix $G$:
\beq\label{data.as.block.of.Gram}
	G = P^TP = \left( \begin{array}{c|c}  G_{\mathrm{st}} & \mathcal{D} \\ \hline  \mathcal{D}^T & G_{\mathrm{m}}   \end{array} \right).
\eeq
Guided by Occam's razor, we described in~\cite{Stark2012GramEstimations}, how $G$ can be estimated via the convex relaxation of a rank minimization problem. In~\cite{Stark2012RigidityConsiderations} we have studied the extent to which the state-measurement Gram matrix is evidence based, i.e., to which extent $G$ is determined by the measurement data $\mathcal{D}$ alone. The state-measurement Gram matrix determines the density matrices and the POVM elements uniquely up to simultaneous rotations (with respect to the Hilbert-Schmidt inner product) of all these Hermitian matrices. 

However, in some applications knowing $G$ might not be sufficient and explicit realizations of the states and the POVM elements in terms of quantum mechanically valid matrices (positive semidefinite) are desired. Finding explicit, quantum mechanically valid, representations is the goal of this paper. For this purpose, we have designed a heuristic algorithm which takes the state-measurement Gram matrix $G$ as input, and returns proper states and POVMs. From a geometric perspective, our task is to fit $W + VK$ (i.e., total number of states plus total number of POVM elements) matrices into the cone of positive semidefinite matrices such that their pairwise inner products satisfy the constraints described by $G$.
%\subsection{General realizations}
%\subsection{Positive semidefinite realizations}
Thus, the wanted states-measurement matrix $P$ can be regarded as the solution to the optimization problem
\beq\label{realization.task.as.optimization.prob}
\begin{aligned}
	 &\text{argmin} & & \| P^TP - G \|_{2}^2 \\
	& \text{subject to}
	& & \text{Columns of $P$ carry valid}\\
	 & & & \text{states and measurements.}
\end{aligned}
\eeq
Here, the constraints refer to the usual demands $\rho_{w} \geq 0$, $\tr(\rho_{w}) = 1$ for all states, and $E_{vk} \geq 0$, $\sum_{k} E_{vk} = \id$ for all POVM elements. Unfortunately, this is not a convex optimization problem. (Analogous to $\min_{x \in \AR} | x^2 -1 |$.) It might thus seem that \eqref{realization.task.as.optimization.prob} is intractable. 
However, in contrast to many other non-convex optimization problems, we have the advantage that we easily can verify when we have reached a global minimum, since the optimal $P$ satisfies $P^TP = G$. This holds true as long as $G$ is a Gram matrix that can approximately be reconstructed by a quantum model. It should be stressed that the  algorithm cannot be treated as a `black box solution' to the task of finding proper states and POVMs. 
We cannot exclude that its application may require modifications of the suggested parameter values in the algorithm. If the algorithm does not converge quickly enough to a desired realization, the evolution of the computation needs to be analyzed to adjust these parameters.  This applies mainly to the parameters dictating the switching from one phase of the algorithm (REGULAR, SELECTION\_OF\_FASTEST, and PARTIAL; see below) to another phase.

%%%%%
\section{\label{Sec:algorithm}The algorithm}

Here we present the an algorithm to solve~\eqref{realization.task.as.optimization.prob}. The main idea is to iteratively solve sequences of convex optimization problems. 
To this end we first observe that although non-convex, problem~\eqref{realization.task.as.optimization.prob} has a lot of structure. When the first factor is replaced by a constant matrix, $P^T \mapsto P_{0}^T$, we arrive at a family of independent least squares problems, since
\beq
\label{non.convex.problem}
	\| P_{0}^TP - G \|_{2}^2 = \sum_{j} \| P_{0}^T P(:,j)  -  G(:,j) \|_{2}^2.
\eeq
Here, $P(:,j)$ and $G(:,j)$ refer to the $j$-th columns of $P$ and $G$ respectively. Each summand of the RHS of (\eqref{non.convex.problem}) corresponds to a convex optimization problem, which thus can be solved efficiently and reliably. This observation leads to Algorithm~\ref{draft.SESAM.realizations}; see Listing~\ref{REGULAR} for the description of the routine 'REGULAR'. There, $\mathrm{mat}(\cdot)$ refers to the inverse of the transition $\rho \mapsto \vec{\rho}$.

\begin{algorithm}
\caption{Draft of SESAM\_realizations}
\label{draft.SESAM.realizations}
\begin{algorithmic}[1]

\Require Gram matrix $G$, Hilbert space dimension $d$, error threshold

\State Initialize $P$ (cf.~Eq.~\eqref{reminder.def.matrix.carrying.st.and.m}) with random density matrices and POVM elements.

\State error := $\infty$

\While{error $\geq$ threshold}
	
	\For{ind = 1:size(P,2)}
		  
		  \State call REGULAR
			 
	\EndFor		

	\State error = $\| P^TP - G \|_{2} / \| G \|_{2}$
	
\EndWhile

\State Density matrices and POVM elements correspond to columns of $P$; cf.~Eq.~\eqref{reminder.def.matrix.carrying.st.and.m}.

\end{algorithmic}
\end{algorithm}

\begin{algorithm}
\floatname{algorithm}{Listing}
\caption{REGULAR}
\label{REGULAR}
\begin{algorithmic}[0]
		
		\State Determine P(:,ind) via
		\beq\begin{aligned} \nn
	 	&\text{argmin}_{\vec{v}} & & \| P^T\vec{v} - G_{:,\text{ind}} \|_{2} \\
		& \text{subject to}
		& & \text{mat($\vec{v}$) positive semidefinite}
		\end{aligned}\eeq
		 
\end{algorithmic}
\end{algorithm}

By explicit tests of Algorithm~\ref{draft.SESAM.realizations} on simple models, one observes that it generally appears to work well on single qubits, or when we remove the constraints that mat($P(:,j)$) must be positive semidefinite for all $j$. 
However, for estimations in $(d > 2)$-dimensional quantum systems, the Algorithm~\ref{draft.SESAM.realizations} appears to easily get stuck away from a global minimum. We suspect that this is connected to the somewhat `edgy' nature of the cone of positive matrices over $\mathbb{C}^{d}$ ($d > 2$). 

The observation that the vectors $\vec{v}$ in subroutine REGULAR get stuck reflects that the demands imposed on $\vec{v}$ are too restrictive if $\mathrm{mat}(\vec{v})$ is cornered in an unfortunate part of the cone of positive semidefinite matrices. This brings us to the introduction of subroutine PARTIAL (see Algorithm~\ref{SESAM.realizations}, line~\ref{place.where.PARTIAL.appears} and Listing~\ref{PARTIAL}).

\begin{algorithm}
\caption{SESAM\_realizations}
\label{SESAM.realizations}
\begin{algorithmic}[1]

\Require Gram matrix $G$, Hilbert space dimension $d$, error threshold

\State Initialize $P$ (cf.~Eq.~\eqref{reminder.def.matrix.carrying.st.and.m}) with random density matrices and POVM elements.

\State mode := 'PARTIAL'

%\State shifts := ones(1,$2*$size($P$,2))

\State error := $\infty$

\State Use $G$ to compute euclidean distances between columns of $P$. Store all relative distances within a matrix $D_{\mathrm{rel}}$; $D_{\mathrm{rel}}(i,j) := \sqrt{G_{ii} - 2 G_{ij} + G_{jj}}$ \label{SESAM.computation.of.relative.dist}

\While{error $\geq$ threshold}
	
	\For{ind = 1:size(P,2)}
	 	\State old\_vect := P(:,ind)

			\Switch{mode}
		    		\Case{'PARTIAL'}
						\State call PARTIAL \label{place.where.PARTIAL.appears}
						\State break
				\EndCase
				  \Case{'SELECTION\_OF\_FASTEST'}
						\State call SELECTION\_OF\_FASTEST
						\State break
				\EndCase
				  \Case{'REGULAR'}
						\State call REGULAR
						\State break
				\EndCase
			\EndSwitch
		
		\State call monitoring
	 
	\EndFor		

	\State call initialization\_of\_switching\_parameters
	\State call switching\_decision
\EndWhile

\State Density matrices and POVM elements correspond to columns of $P$; cf.~Eq.~\eqref{reminder.def.matrix.carrying.st.and.m}.

\end{algorithmic}
\end{algorithm}

\begin{algorithm}
\floatname{algorithm}{Listing}
\caption{PARTIAL}
\label{PARTIAL}
\begin{algorithmic}[0]
	 	
		\State Use $D_{\mathrm{rel}}$ (cf. line~\ref{SESAM.computation.of.relative.dist} in Algorithm~\ref{SESAM.realizations}) to compute the index set ind\_NH containing the indices of the $d^2$ columns of $P$  which are closest to $P(:,\text{ind})$. Then, determine P(:,ind) via
		
		\beq\begin{aligned} \nn
	 	&\text{argmin}_{\vec{v}} & & \| [P^T\vec{v}]_{\text{ind\_NH}} - G_{\text{ind\_NH,ind}} \|_{2} \\
		& \text{subject to}
		& & \text{mat($\vec{v}$) positive semidefinite}
		\end{aligned}\eeq
		 
\end{algorithmic}
\end{algorithm}

The purpose of running PARTIAL is to remove some of the demands specified in terms of $G(:,j)$ in the objective function in REGULAR. More precisely, PARTIAL does not try to align the dynamic $P$-column $\vec{v}$ as well as possible with respect to all states and measurements. Instead it only tries to align $\vec{v}$  as well as possible with the vectors of its local neighborhood. Thus, PARTIAL aims at restoring local neighborhoods while REGULAR tries to directly arrange the whole family of states and measurements. In Listing~\ref{PARTIAL} we chose the minimal choice $d^2$ for the size of the local neighborhood. If we selected fewer vectors, the result of each execution of PARTIAL would not be unique, since the vectors $P(:,j)$ live in a $d^2$-dimensional space (recall that the matrices $\mathrm{mat}(P(:,j))$ are forced to be Hermitian). 

To run PARTIAL, we need to determine the local neighborhood of each column $P(:,j)$ by selecting the $d^2$ nearest neighbors (with respect to Hilbert-Schmidt distance) of $P(:,j)$. The necessary knowledge about all the pairwise relative distances $D_{\mathrm{rel}}(i,j)$ between columns $i$ and $j$ of $P$ can be computed directly via the Gram matrix $G$:
\beq
\begin{split}
	D_{\mathrm{rel}}(i,j)^2 &= \sp{P(:,i) - P(:,j)}{P(:,i) - P(:,j)} \\
					        &= G_{ii} - 2 G_{ij} + G_{jj}.
\end{split}
\eeq
This calculation is performed on line~\ref{SESAM.computation.of.relative.dist} of Algorithm~\ref{SESAM.realizations}. Looking at Algorithm~\ref{SESAM.realizations} we recognize that---for the moment we ignore the routine SELECTION\_OF\_FASTEST---in some iterations of the while-loop we run REGULAR and in some other iterations we run PARTIAL. Thus, we need to prescribe which routine to select. This is specified in Listing~\ref{switching.decision} but we postpone the explanation of the switching criteria for later. 

In summary we have arrived at an Algorithm that switches back and forth between REGULAR and PARTIAL. PARTIAL tries to restore local neighborhoods while REGULAR tries to find the global, complete configuration. REGULAR succeeds if it starts with an initial condition that is close to a solution. This might be achieved via an educated guess. However, even if PARTIAL manages to arrange all the vectors such that supposedly near vectors are near, we have not necessarily constructed a good starting point for REGULAR. This is due to the complicated geometry of the cone of positive semidefinite matrices in case of Hilbert space dimensions $d > 2$. The edged shape of the cone of positive semidefinite matrices makes it necessary to rotate the complete collection of states-measurement vectors as a whole so that the edges of the contour of the states-measurement configuration fits into the edges of the cone of positive semidefinite matrices. Since we only allow one vector to move in each iteration (that is the precondition for using convex optimization in each step) this becomes a tiresome business because the dynamic vector is held back by all the static vectors due to the interactions described by $G(:,j)$. The dynamics of the whole states-measurement configuration thus resembles the movement of a worm: each segment of the worm corresponds to one column $P(:,j)$. We need the worm to travel from $A$ to $B$. Thus, each segment of the worm has to travel from $A$ to $B$. In each time step, only one segment of the worm is dynamic and wants to reach $B$ but its movement is highly constrained by its bonding to the static segments. Although this analogy is admittedly a little far fetched, it nevertheless provides an intuition for the dynamics of the bunch of state-measurement vectors. Speeding up the motion of the states-measurements worm is the main purpose of the remaining routine.

The routine SELECTION\_OF\_FASTEST, see Listing~\ref{SELECTION.of.FASTEST}, takes into account that some worm segments move faster than others, and it operates by ignoring the slowest moving segments. Thus, instead of letting the worm being slowed down by its small segments, we ignore the slow segments. Consequently, during SELECTION\_OF\_FASTEST, the evolution of the columns $P(:,j)$ only tries to align the dynamic vector with respect to the fast moving components of the state-measurement bunch. 

\begin{algorithm}
\floatname{algorithm}{Listing}
\caption{SELECTION\_OF\_FASTEST}
\label{SELECTION.of.FASTEST}
\begin{algorithmic}[0]
	 	
		\State Use shifts (cf. line~\ref{update.of.shifts} in Listing~\ref{monitoring}) to determine the index set ind\_fast containing the indices of the $d^2$ columns of $P$ which have moved the fastest in the preceding round. Then, determine P(:,ind) via
		\beq\begin{aligned} \nn
	 	&\text{argmin}_{\vec{v}} & & \| [P^T\vec{v}]_{\text{ind\_fast}} - G_{\text{ind\_fast,ind}} \|_{2} \\
		& \text{subject to}
		& & \text{mat($\vec{v}$) positive semidefinite}
		\end{aligned}\eeq

\end{algorithmic}
\end{algorithm}

%%%%%
\section{Switching subroutines}\label{Sec:switching}

In the previous section we described the main parts of the algorithm. Here we describe the criteria for the switching between these subroutines. This is specified in Listing~\ref{switching.decision}. To detect the right moment to switch from one routine to the other, we require some basic monitoring of the computation. The monitoring in Listing~\ref{monitoring} records a history of how fast the individual vectors move ($\rightarrow$ shifts), a history of the temporary error $\| P^TP - G \|_{2} / \| G \|_{2}$ ($\rightarrow$ error\_history), and the current trend of the error ($\rightarrow$ trend). Moreover, we store the quantities zenith and counter\_no\_zenith. The variable zenith stores the largest error that has been encountered until the last switching. The quantity counter\_no\_zenith counts for how many iterations the procedure has not updated the value of zenith. 

Listing~\ref{initialization.of.switching.parameters} initializes some parameters necessary for switching\_decision. We have chosen random initializations to break periodicities (we observed that in some situations the complete states-measurement vectors undergoes periodic movements).

\begin{algorithm}
\floatname{algorithm}{Listing}
\caption{monitoring}
\label{monitoring}
\begin{algorithmic}[1]

	\State shift $= \| \text{old\_vect} - P(:,\text{ind}) \|_{2} / \| P(:,\text{ind}) \|_{2}$

	 \State shifts = [shifts shift] \label{update.of.shifts} 
	 
	 \State error = $\| P^TP - G \|_{2} / \| G \|_{2}$
	 
	 \State error\_history = [raw\_history error]
	 
	  \State trend = (error - error\_history(end-size($P,2$)+1) / error\_history(end-size($P,2$+1) 
	 
	 \If{error $>$ zenith}
	 	\State zenith = error
		\State counter\_for\_no\_zenith = 0
	\Else
		\State counter\_no\_zenith = counter\_no\_zenith + 1
	\EndIf	

\end{algorithmic}
\end{algorithm}

We have chosen to switch between the different routines according to the following sequential ordering:
\[\begin{array}[c]{ccl}
	\mathrm{REGULAR}&\stackrel{\circled{1}}{\longrightarrow}&\mathrm{PARTIAL}\\
	&\circled{3} \nwarrow&\downarrow \circled{2}\\
	&&\mathrm{SELECTION\_OF\_FASTEST}
\end{array}\]

$\circled{1}$ We decide to leave REGULAR whenever staying in this phase does not lead to meaningful improvements. The variable trend is updated in Listing~\ref{monitoring}. The threshold switch\_REG\_to\_PARTIAL is randomly updated in Listing~\ref{initialization.of.switching.parameters}.

$\circled{2}$ This transition is enforced under two circumstances. In the first case, we switch whenever $\| \text{shifts(end-$2*$size($P,2$)+1:end)} \|_{\infty} \leq$ 0.002, i.e., we switch whenever PARTIAL gets stuck. The the second case occurs when the procedure has not encountered a new zenith in no\_zenith\_threshold-many iterations. This rule is motivated as follows: Recall that we start PARTIAL after REGULAR got stuck in a local minimum (eventually just a plateau). We apply PARTIAL with the intention that it should lift the states-measurement configuration out of the local minimum. To do so, we have to cross an `error-barrier'. If the procedure has not reached a new zenith in no\_zenith\_threshold-many iterations, then we take this as an indication that the states-measurement configuration may have crossed a barrier, and we could check whether or not the bottom of the current valley coincides with the global minimum $\| P^TP - G \|_{2} = 0$.

$\circled{3}$ Before exploring the valley with REGULAR, we first apply SELECTION\_OF\_FASTEST, since we do anticipate the `worm-like' dynamics described in the previous section. This phase is left when it yields no more improvements. Usually, this happens rather quickly because typically, SELECTION\_OF\_FASTEST leads to chaotic movements soon after its initiation.

\begin{algorithm}
\floatname{algorithm}{Listing}
\caption{initialization\_of\_switching\_parameters}
\label{initialization.of.switching.parameters}
\begin{algorithmic}[1]

	\State Throw a random coin; prob('heads') = 0.7
	 
	 \If{coin='heads'}
	 	\State switch\_FAST\_to\_REG = -0.08
	 \Else
	 	\State switch\_FAST\_to\_REG = -0.05
	 \EndIf		
	 
	 \State Throw a random coin; prob('heads') = 0.7
	
	 \If{coin='heads'}
	 	\State switch\_REG\_to\_PARTIAL = -0.01
	 \Else
	 	\State switch\_REG\_to\_PARTIAL = -0.02
	 \EndIf
	 
	 \State Throw a random coin; prob('heads') = 0.7
	
	 \If{coin='heads'}
	 	\State no\_zenith\_threshold = $3 * \mathrm{size}(P,2)$
	 \Else
	 	\State no\_zenith\_threshold = $7 * \mathrm{size}(P,2)$
	 \EndIf
	
\end{algorithmic}
\end{algorithm}

\begin{algorithm}
\floatname{algorithm}{Listing}
\caption{switching\_decision}
\label{switching.decision}
\begin{algorithmic}[1]
  \Switch{mode}
    \Case{'PARTIAL'}
     		 \If{ $\| \text{shifts(end-$2*$size($P,2$)+1:end)} \|_{\infty} \leq$ 0.002 }
		 	\State mode = 'SELECTION\_OF\_FASTEST'
			\State zenith = -1
			\State break
		 \EndIf
		 
		 \If{ counter\_for\_no\_zenith $>$ no\_zenith\_threshold }
		 	\State mode = 'SELECTION\_OF\_FASTEST'
			\State zenith = -1
			   \State break
		 \EndIf
    \EndCase

    \Case{'SELECTION\_OF\_FASTEST'}
      		   \If{ trend $\geq$ switch\_FAST\_to\_REG }
		 	\State mode = 'REGULAR'
			\State zenith = -1
			   \State break
		 \EndIf
    \EndCase
    
    \Case{'REGULAR'}
      		  \If{ trend $\geq$ switch\_REG\_to\_PARTIAL }
		 	\State mode = 'PARTIAL'
			\State zenith = -1
			   \State break
		 \EndIf
    \EndCase
  \EndSwitch
\end{algorithmic}
\end{algorithm}

%%%%%
\section{Examples}\label{Sec:examples}

The goal of this section is to present our experiences when running the proposed algorithm~\ref{SESAM.realizations} in different situations. We are considering 2- to 4-dimensional systems with different levels of randomness in the states and the POVM elements. We cover the following three cases:
\begin{itemize}
\item \emph{Scenario `pure'.} We are sampling pure states and projective measurements uniformly from the Haar measure.

\item \emph{Scenario `partly mixed'.} First, we uniformly sample $\eta_{w}$ from $[\eta_{\mathrm{lb}}, \eta_{\mathrm{ub}}]$ for each state `$w$', and $\mu_{vk}$ from $[\mu_{\mathrm{lb}}, \mu_{\mathrm{ub}}]$ for each POVM element `$(vk)$'. We chose $\eta_{\mathrm{lb}} = \mu_{\mathrm{lb}} = 0.6$ and $\eta_{\mathrm{ub}} = \mu_{\mathrm{ub}} = 0.8$. To generate the states, we sample separately for each state a unitary $U_{w}$ from the Haar measure, and define 
\[
	\rho_{w} := U_{w} \, \mathrm{diag}(\eta_{w}, 1-\eta_{w}, 0, ..., 0) \, U_{w}^*.
\]
To generate the POVM elements, we set $P_{vk}$, such that 
\[
	\bigl( P_{vk} \bigr)_{ij} := \mu_{vk} \delta_{ij} \delta_{ik}.
\]
Then, for $k=1,...,(K-1)$,
\[
	E_{vk} :=  U_{vk} \, P_{vk} \, U_{vk}^*, 
\]
and
\[
	E_{vK} := \id - \sum_{k=1}^{K-1} E_{vk}.
\]
Here, all the unitary matrices are again sampled according to the Haar measure.

\item \emph{Scenario `purified'.} All POVM effects on $\HR$ associated to the $K$ possible outcomes are of the form $\{ \bra{k} U \ket{a} \}_{k=1}^K$ where $U$ is a unitary on $\HR \otimes \HR_{\mathrm{anc}}$, $\HR_{\mathrm{anc}}$ a $K$-dimensional ancilla system, $\{ \ket{k} \}_{k=1}^K$ is an orthonormal basis in $\HR_{\mathrm{anc}}$, and $\ket{a}$ is an arbitrary state in $\HR_{\mathrm{anc}}$. Moreover, for all choices of $U$, the definitions
\[
	M_{k} := \bra{k} U \ket{a}
\]
yield valid effects and therefore, $E_{k} := M_{k}^* M_{k}$ defines a POVM. In the present scenario `purified', we construct the POVM `$v$' by sampling $U_{v}$ on $\HR \otimes \HR_{\mathrm{anc}}$ with respect to the Haar measure, by defining $M_{vk} := \bra{k} U_{v} \ket{1}$, and by setting $E_{vk} := M_{vk}^* M_{vk}$. To sample the states, we set
\[
	\rho_{w}^{(0)} := \mathrm{diag}(p_{w1}, ..., p_{wd}) 
\]
with $p_{w\alpha}$ sampled uniformly from $[0,1]$. Then,
\[
	\rho_{w} := \frac{U_{w} \, \rho_{w}^{(0)} \, U_{w}^*}{\tr(\rho_{w}^{(0)})}
\]
with $U_{w}$ sampled from the Haar measure.

\end{itemize}

\begin{figure}[htb]
\centering
\includegraphics[width=0.38\columnwidth]{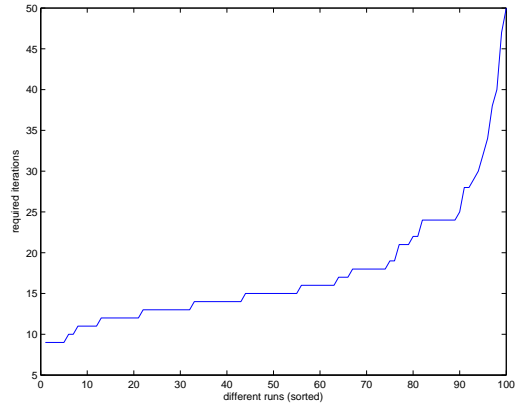}
\caption{Scenario `pure', $(d,W,V,K) = (2,8,4,2)$.}
\label{2D.pure}
\end{figure}

\begin{figure}[htb]
\centering
\includegraphics[width=0.38\columnwidth]{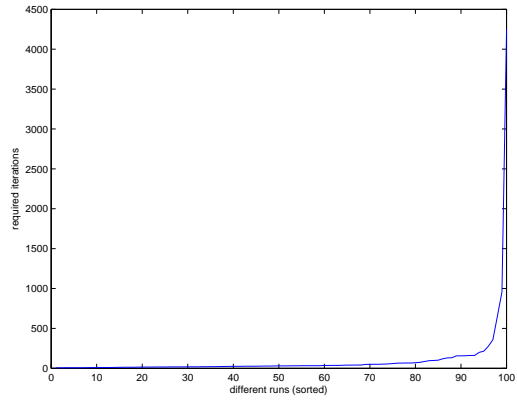}
\caption{Scenario `partly mixed', $(d,W,V,K) = (2,8,4,2)$.}
\label{2D.partly.mixed}
\end{figure}

\begin{figure}[htb]
\centering
\includegraphics[width=0.38\columnwidth]{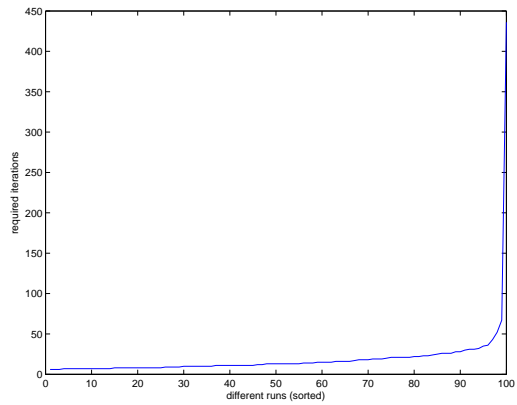}
\caption{Scenario `purified', $(d,W,V,K) = (2,8,4,2)$.}
\label{2D.purified}
\end{figure}

\begin{figure}[htb]
\centering
\includegraphics[width=0.38\columnwidth]{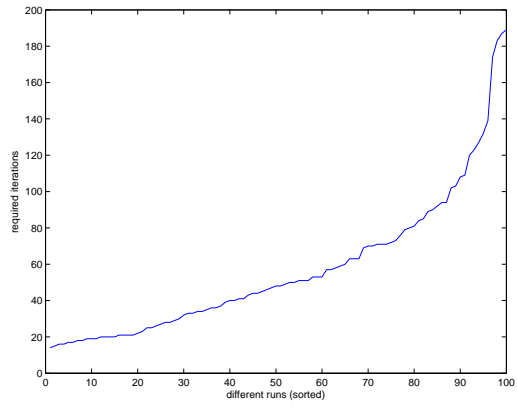}
\caption{Scenario `pure', $(d,W,V,K) = (3,27,9,3)$.}
\label{3D.pure}
\end{figure}

\begin{figure}[htb]
\centering
\includegraphics[width=0.38\columnwidth]{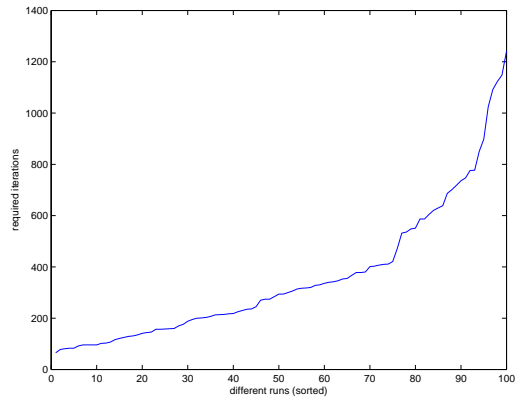}
\caption{Scenario `partly mixed', $(d,W,V,K) = (3,27,9,3)$.}
\label{3D.partily.mixed}
\end{figure}

\begin{figure}[htb]
\centering
\includegraphics[width=0.38\columnwidth]{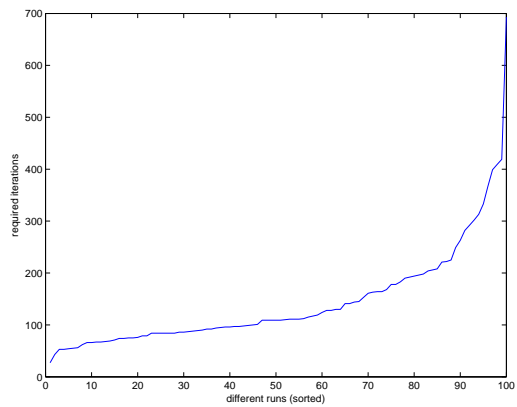}
\caption{Scenario `purified', $(d,W,V,K) = (3,27,9,3)$.}
\label{3D.purified}
\end{figure}

\begin{figure}[htb]
\centering
\includegraphics[width=0.38\columnwidth]{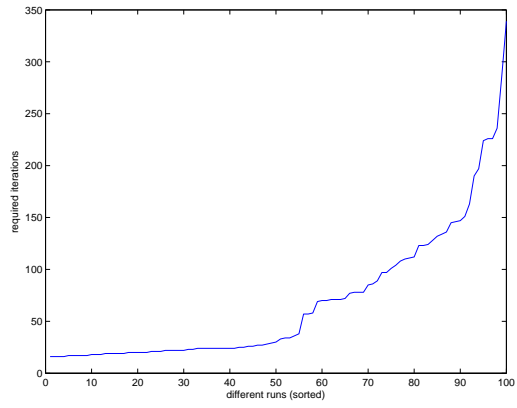}
\caption{Scenario `pure', $(d,W,V,K) = (4,56,9,4)$.}
\label{4D.pure}
\end{figure}

\begin{figure}[htb]
\centering
\includegraphics[width=0.38\columnwidth]{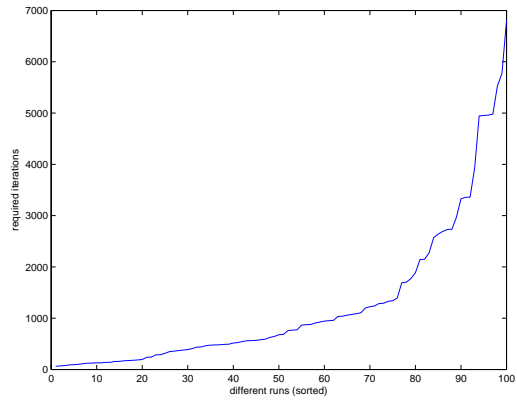}
\caption{Scenario `partly mixed', $(d,W,V,K) = (4,56,9,4)$.}
\label{4D.partly.mixed}
\end{figure}

\begin{figure}[htb]
\centering
\includegraphics[width=0.38\columnwidth]{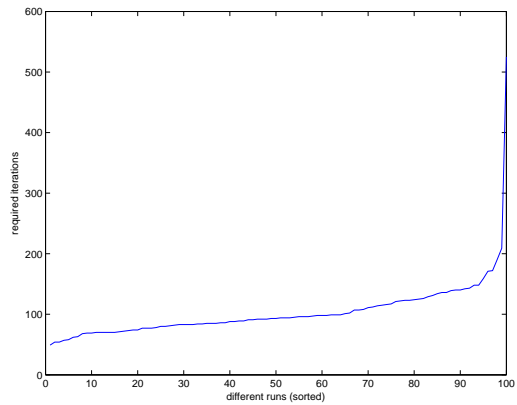}
\caption{Scenario `purified', $(d,W,V,K) = (4,56,9,4)$.}
\label{4D.purified}
\end{figure}

We have run each of the scenarios `pure', `partly mixed', and `purified' 100 times for 2-, 3-, and 4-dimensional systems. The stopping criteria was
\beq
	\max_{ij}\{ | \tr(\rho_{i} E_{j}) - G_{ij}| \} \leq 10^{-2}, \, \forall i,j.
\eeq
Each run converged successfully. However, we observe that some cases required considerably more iterations (i.e., calls of the subroutines PARTIAL, SELECTION\_OF\_FASTEST, and REGULAR) than others. This is visible in Fig.~\ref{2D.pure} to Fig.~\ref{4D.purified}. Table~\ref{numerical.experiments.SESAM} summarizes our findings. Calculations have been performed using CVX~\cite{cvx,gb08} calling SeDuMi~\cite{Sturm1999}.

\begin{table}[htdp]
\caption{Numerical findings}
\begin{center}
\begin{tabular}{|l|c|c|c|c|c|c|}
	\hline Scenario    		 	& $d$& $W$ 	& $V$& $K$ & successes 	& failures\\ 
	\hline\hline `pure'  		 	& 2	& 8 		& 4 	& 2	& 100 		& 0\\ 
	\hline         `partly mixed'	 	& 2	& 8 		& 4 	& 2	& 100		& 0\\
	\hline         `purified' 		 	& 2	& 8 		& 4	& 2	& 100		& 0\\
	 \hline 	`pure' 		    	& 3	& 27		& 9	& 3	& 100 		& 0\\
	  \hline 	`partly mixed'  	 	& 3	& 27		& 9	& 3	& 100 		& 0\\ 
	  \hline 	`purified' 	 	 	& 3	& 27		& 9	& 3	& 100 		& 0\\ 
	  \hline 	`pure' 		 	& 4	& 56 	& 9	& 4	& 100 		& 0\\ 
	  \hline 	`partly mixed'	 	& 4	& 56 	& 9	& 4	& 100 		& 0\\ 
	  \hline 	`purified'  			& 4	& 56 	& 9	& 4	& 100		& 0 \\ 
	\hline
\end{tabular}
\end{center}
\label{numerical.experiments.SESAM}
\end{table}%

%- Discuss the outliners, and suggest chances in the parameters to speed up their convergence. For instance, small set of vectors which are approximately parallel might need more tolerance in the REG phase to reach convergence.

%%%%%
\section{Conclusion}

We have introduced a heuristic algorithm to generate proper density operators and POVMs from the knowledge of the Gram matrix of a collection of a priori unknown states and measurements. In other words, based only on the knowledge of the (Hilbert-Schmidt) inner products between states and POVM elements, between states and states, and between POVM elements and POVM elements, this technique delivers proper density operators and POVM elements, i.e., operators that are positive semidefinite and satisfy the standard normalization conditions. We do not know to what extent this computation of operators in the positive cone, consistent with the acquired data, is a computationally hard problem. 

This paper concludes a proof of principle that it is possible to simultaneously estimate the dimension, states, and measurements in the asymptotic regime. This program was introduced in~\cite{Stark2012GramEstimations}, where we showed how to estimate the Gram matrix $G$ associated to the involved states and POVM elements. In \cite{Stark2012RigidityConsiderations} we focused on the freedom in choosing $G$, and introduced a method to test whether or not specific assumptions on the Gram matrix can guarantee its uniqueness. By the extraction of states and POVMs presented here, we have thus completed the task of simultaneous estimation of dimension, states and POVMs.

However, several questions remain concerning all the steps of the estimation procedure, ranging from scenarios involving non-asymptotic sampling (leading to statistical fluctuations in the Gram matrix $G$) to efficient, stable, and practically applicable numerical procedures. 

%Furthermore, we have throughout this analysis assumed an asymptotic setting, i.e., we have assumed access to effectively infinite data sets. A relevant question is thus the stability of this estimation procedure under statistical errors, i.e., in the (more realistic) setting of finite sampling.

%- Stress that the present work is a proof of principle.\\
%
%- Need for non-asymptotic treatments.\\
%
%- QM as general data analysis framework.\\
%
%- Interpretation of QM.\\
%
%- Uniqueness still incomplete. Mention Cucuringu's algebraic geometry open questions and that the presented local rigidity approach does not take into account the cone constraints and the linear constraint given via the trace one condition for the states and the summation to the identity constraint for the POVM elements.\\
%
%- What happens if you use measurements as preparation devices? Can you determine diagonal entries in some circumstances? Only if measurements are pure and non-degenerate?
%
%
%
%- Schmaler Strauss im Kegel: nicht lokal rigid. Dennoch wird lokal rigid durch generische Situationen festgelegt. Wie passt dies zusammen?

%%%%%
\section*{Acknowledgements}

I wish to express my gratitude to Johan {\AA}berg for helping me to finish this paper. Moreover, I want to thank Matthias Baur, Matthias Christandl, Dejan Dukaric, Fr\'ed\'eric Dupuis, Philippe Faist, David Gross, Patrick Pletscher, Renato Renner, Lars Steffen, L\'idia del Rio, Sharon Wulff, and M\'ario Ziman for interesting discussions. I acknowledge support from the Swiss National Science Foundation through the National Centre of Competence in Research ``Quantum Science and Technology''.

%\bibliographystyle{unsrt} %unsrt used to order references according to their appearance in the text; otherwise: plain
%\bibliography{SESAM_finding_explicit_realizations}

%%%%%%%%%%%%%%%%%
%%%%%%%%%%%%%%%%%
%%%%%%%%%%%%%%%%%
%%%%%%%%%%%%%%%%%

\clearpage

\widetext

%%%% 2D

%%%% 3D

%%%% 4D

\end{document}